\documentclass[12pt,twoside]{article}

\usepackage{amsmath}
\input BoxedEPS
\SetRokickiEPSFSpecial  
\HideDisplacementBoxes

\makeatletter
\long\def\@makecaption#1#2{%
  \vskip\abovecaptionskip
  \sbox\@tempboxa{\small #1: #2}%
  \ifdim \wd\@tempboxa >\hsize
   \small #1: #2\par
  \else
    \global \@minipagefalse
    \hb@xt@\hsize{\hfil\box\@tempboxa\hfil}%
  \fi
  \vskip\belowcaptionskip}
\makeatother

\newcommand{\bra}[1]{\bigl\langle#1\bigl|}
\newcommand{\ket}[1]{\bigr|#1\bigr\rangle}
\newcommand{\znn}{z_1,\ldots,z_n}
\newcommand{\zn}{z_1,\ldots,z_n}

\newcommand{\zsn}{z_1 + z_2 +\cdots + z_n}

\newcommand{\Ht}{H(t)}

\newcommand{\ps}[1]{\ket{\psi(#1)}}
\newcommand{\rd}[1]{\mathop{\mathrm{d}#1}}

\newcommand{\fract}[2]{{\textstyle\frac{#1}{#2}}}

\newcommand{\Bd}{Boltz\-mann distribution}

\newcommand{\aea}{adiabatic evolution algorithm}
\newcommand{\qa}{quantum algorithm}
\newcommand{\qaa}{quantum adiabatic algorithm}
\newcommand{\qae}{quantum adiabatic evolution}
\newcommand{\qaea}{quantum adiabatic evolution algorithm}

\newcommand{\sat}{{\sc sat}}

\newcommand{\numeq}[2]{\begin{equation}
#2
\label{#1}
\end{equation}}
\newcommand{\refeq}[1]{(\ref{#1})}

\newcommand{\citer}[1]{~\cite{#1}}

\let\epsilon\varepsilon
\let\phi\varphi
\let\tilde\widetilde

\newcommand{\refFig}[1]{Figure~\ref{#1}}

\begin{document}
\title{Quantum Adiabatic Evolution Algorithms\\ 
 versus Simulated Annealing}
\author{Edward Farhi, Jeffrey
Goldstone\footnote{\tt farhi@mit.edu,
goldston@mit.edu}\\[-.75ex]
\footnotesize\itshape Center for Theoretical Physics, Massachusetts
Institute of Technology, Cambridge, MA 02139
\\ 
Sam Gutmann\footnote{\tt sgutm@neu.edu}\\[-.75ex]
\footnotesize\itshape  Department of Mathematics, Northeastern University,    
 Boston, MA 02115\\[-1.5ex]}
\date{\footnotesize\sf MIT-CTP~\#3228 \qquad  quant-ph/0201031}
\maketitle
\pagestyle{myheadings}
\markboth{E. Farhi, J.  Goldstone,   S. Gutmann}{Quantum Adiabatic Evolution
Algorithms 
 versus Simulated Annealing}
\vspace*{-2.25pc}

\begin {abstract}\noindent
We explain why \qae\ and simulated annealing perform similarly in certain
examples of searching for the minimum of a cost function of $n$ bits. In these
examples each bit is treated symmetrically so the cost function depends only on
the Hamming weight of the $n$ bits. We also give two examples, closely related to
these, where the similarity breaks down in that the \qaa\ succeeds in polynomial
time whereas  simulated annealing requires exponential time.
\end{abstract}

\vspace*{-1pc}

\thispagestyle{empty}

\setcounter{equation}{0}
\subsection{Introduction}
\label{sec1}

Quantum \aea s\citer{r1} are designed to minimize a (classical) cost function whose
domain is the $2^n$ values taken by $n$~bits.  To test the algorithm it is
natural to look at problems where classical local search algorithms, such as
simulated annealing, have difficulty. It is easy to construct examples where both
classical local search and \qae\ require time exponential in~$n$. An example is
the Grover problem, where the function takes the value~0 at a single input and
is~1 on the $2^n -1$ other inputs. A variant of this Grover example can be found
in\citer{r2}.

More interesting examples arise when the cost function is local in the sense that
it can be written as a sum of terms each of which involves only a few bits.
A~3\sat\ example where both  the classical local search and ``standard'' \qaa s
are exponentially slow has recently been given in\citer{r3}.  In
Section~\ref{sec4}
below we  analyze an example that is based on\citer{r3} and
give a quantum mechanical argument for why \qaa s have trouble with these kinds
of examples. 

In Section~\ref{sec5} we give a nonlocal example where simulated annealing
requires exponential time but \qae\ succeeds in polynomial time. In
Section~\ref{sec6} we give a  detailed analysis of a local (2\sat) example that
first appeared in\citer{r1}. Here again a \qaa\ succeeds even though simulated
annealing fails in polynomial time. These examples show that there is no general
theorem that
\qaa s must fail if simulated annealing fails.

\subsection{Quantum adiabatic algorithms}
\label{sec2}

Quantum \aea s offer a general approach to solving combinatorial search
problems on a quantum computer.  A local combinatorial search problem can be
cast as the classical problem of finding the minimum of cost function $h(\zn)$
where each $z_i=0,1$ and 
\numeq{e1}{
 h = \sum_C h_C
} 
and each $h_C$ is a nonnegative integer-valued function that depends only on a
few~$z_i$. A \qaa\ converts this into the problem of producing the ground state
of a quantum Hamiltonian $H_P$ given by 
\numeq{e2}{
H_P \ket{\zn} = 
h(\zn) \ket{\zn}  
}
where $\{\ket{\zn}\}$ is a basis for the $2^n$-dimensional Hilbert space of the
quantum computer. Specifying the algorithm consists of choosing a smoothly
varying Hamiltonian $\tilde H(s)$ for $0\le s\le1$ such that $\tilde H(1)=H_P$
and where $\tilde H(0)$ has a ground state $\ket{\psi_0}$ that is known and
easily constructed. A run time~$T$ must also be specified. The Hamiltonian
that governs the evolution is given by
\numeq{e3}{
\Ht = \tilde H(t/T) 
}
and the state of the quantum computer evolves according to the Schr\"odinger
equation 
\numeq{e4}{
i \frac {\rd{}}{\rd t} \ps t
 = H(t) \ps t
}
with the state at time~$t=0$ given by
\numeq{e5}{
\ps0 = \ket{\psi_0}\ .
}
Under conditions that generally apply in the cases of interest here, the quantum 
adiabatic theorem guarantees that for~$T$ large enough the state of the
quantum computer, $\ps t$  for $0\le t\le T$, will be close to the (instantaneous)
ground state of
$\Ht$, and in particular $\ps T$ will be close to the ground state of $H_P$,
encoding the solution to the problem at hand. When applied to a particular
combinatorial search problem, the \qa\ is considered to be successful if the
required running time grows polynomially in the number of  bits~$n$.

The required running time can be related to the spectrum of $\tilde  H(s)$, in
particular the difference between the two lowest eigenvalues $E_0(s)$ and
$E_1(s)$. The required running time~$T$  must obey
\numeq{e6}{
T \gg \frac{\mathcal E}{\mathrm{gap}^2}
}
where
\numeq{e7}{
\mathrm{gap} = \min_{0\le s\le1} \bigl(E_1(s) - E_0(s)\bigr) 
}
and $\mathcal E$ is less than the largest eigenvalue of $H_P - \tilde H(0)$, always
polynomial in~$n$ in the examples here.

\subsection{Simulated annealing}
\label{sec3}

Simulated annealing is a classical  local search strategy that can be used to find
the global minimum of a function $h(\zn)$ of the form \refeq{e1}.
For any temperature $\tau$, $0\le\tau\le\infty$, define the Boltzmann
distribution by 
\numeq{e8}{
P_\tau(\zn) = c(\tau) \exp \bigl[ -h(\zn)/\tau\bigr]
}
where $c(\tau)$ is the normalizing constant such that
\numeq{e9}{
\sum_{\zn} P_\tau (\zn) = 1\ .
}
For $\tau=\infty$, all $2^n$ strings $\zn$ are equally likely.  For $\tau=0$, the
 \Bd\ concentrates on the global minimum (or minima) of~$n$. 

The idea is to construct a Markov chain that starts in the $\tau=\infty$ \Bd\ and
gradually moves through the \Bd s for decreasing~$\tau$ down to $\tau$
near~0. If the process succeeds, the final distribution will be close to  the  zero 
temperature \Bd, which means that the global minimum of~$h$ has been found
with high probability.  More specifically, choose a sequence of temperatures
$\infty\ge \tau_1\ge \tau_2\ge \cdots \tau_M\ge 0$ at which bit values may
change. The $\tau_i$ can be chosen deterministically or randomly. Also
choose a sequence of bits $i_1,i_2,\ldots,i_M$. Bit $i_k$ is the bit that might be
changed (according to an acceptance rule to be given below  in \refeq{e10}) at
step~$k$.  The sequence $i_1,i_2,\ldots,i_M$ can be deterministic or random. 

The initial string is picked uniformly, i.e., all $2^n$ strings $\zn$ have
probability\linebreak $2^{-n}$. At step~$k$, if the current string is
$z_1,\ldots,z_{i_k},
\ldots, z_n$ we flip bit $i_k$ and move to string 
$z_1,\ldots,(1-z_{i_k}), \ldots, z_n$
with probability
\begin{gather} 
\min\bigl\{1, \exp \bigl( -\Delta h/ \tau_k\bigr)\bigr\} \nonumber\\
\text{where}\quad
\Delta h = h(z_1,\ldots,(1-z_{i_k}), \ldots, z_n) -
h(z_1,\ldots,z_{i_k}, \ldots, z_n)\, . \label{e10}
\end{gather}
(This is the Metropolis rule.) Otherwise the string remains unchanged. Note that bit
$i_k$  is always flipped if the change decreases~$h$, and is flipped with only small
probability if the change would increase $h$ by a large amount relative to the
current temperature.

The acceptance rule~\refeq{e10} guarantees that if \emph{many} transitions
 are made at a \emph{single} temperature~$\tau$,  the
distribution of the string will converge to the \Bd\ at temperature~$\tau$. If the
sequence $\tau_1,\tau_2,\ldots,\tau_M$ decreases slowly enough, the
distribution, which starts in the $\tau=\infty$ \Bd, will at step~$k$ be close to
the \Bd\ for temperature~$\tau_k$. The question is how many transitions must
be made in order to stay near the \Bd s as the temperature is lowered down to
near~0. If the total number of steps required grows only as a polynomial
in~$n$, we say that the simulated annealing algorithm succeeds.

\subsection{Symmetrized cost functions}
\label{sec4}
 
\begin{figure}[tb]
\centerline{\BoxedEPSF{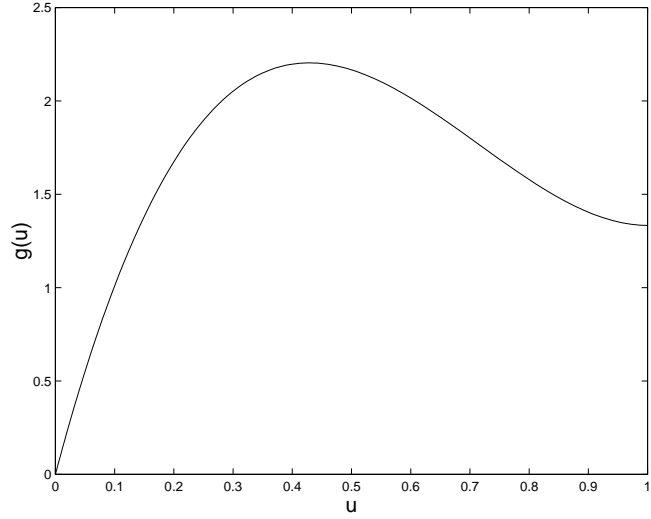 scaled 500}}
\caption{The leading term in the scaled cost as a function of the scaled Hamming
weight $u=w/n$.}
\label{f1}
\end{figure}

In this section we discuss cost functions of the  form \refeq{e1} that are functions
only of the Hamming weight $w=\zsn$ as a result of all bits being treated
symmetrically. As an example, let
$h_3$ depend on  three bits $z,z',\ \text{and}\ z''$  with the form
\numeq{e11}{
h_3 (z ,z',z'') = \begin{cases}
0 & z + z' + z'' = 0\\
q & z + z' + z'' = 1\\
1 & z + z' + z'' = 2\\
1 & z + z' + z'' = 3 
\end{cases}
}
where $q$ is an integer greater than or equal to~3. Now let
\numeq{e12}{
h(\znn) = \sum_{i<j<k} h_3(z_i,z_j,z_k)
} 
which gives 
\begin{align}
h(w) &\equiv h(\znn) \nonumber\\
 &= \fract q2 w(n-w)(n-w-1) + \fract12  w(w-1)(n-w) + \fract16 w(w-1)(w-2)\ .
\label{e13}
\end{align}
We can also write
\numeq{e14}{
h(w) = \Bigl(\frac n2\Bigr)^3 g(w/n) + O(n^2)
}
where
\numeq{e15}{
g(u) = 4 qu(1-u)^2 + 4u^2(1-u) + \fract43 u^3\ .
}
For $n$ large the form of $g(u)$ determines the behavior of the  classical and
quantum algorithms. In \refFig{f1} we plot $g(u)$ for $q=3$. Note that the
global  minimum of $g$ is at $u=0$ corresponding to the unique input
$\znn=0,\dots,0$. The derivative $g'(1/2)$ is negative and there is a local
minimum at $u=1$ corresponding to $\znn=1,\dots,1$.

We now look at the quantum algorithm applied to this problem. The cost
function~\refeq{e13} corresponds to the Hamiltonian 
\begin{multline} \label{e16}
H_P = \frac q2 \Bigl(\frac n2 - S_z\Bigr)\Bigl(\frac n2 + S_z\Bigr)\Bigl(\frac
n2+S_z -1\Bigr)\\
+ \frac12 \Bigl(\frac n2 - S_z\Bigr)\Bigl(\frac n2 - S_z -1\Bigr)\Bigl(\frac
n2+S_z\Bigr)\\
+ \frac16 \Bigl(\frac n2 - S_z\Bigr)\Bigl(\frac n2 - S_z -1\Bigr)\Bigl(\frac
n2-S_z-2\Bigr)
\end{multline}
where
\numeq{e17}{
S_z \ket{\znn} = \Bigl(\frac n2 - w\Bigr) \ket{\znn}\ .
}
We choose for the initial Hamiltonian (following\citer{r1}, formula 2.28)
\numeq{e18}{
\tilde H(0) = \Bigl(\begin{matrix} n-1\\2\end{matrix}\Bigr)\Bigl(\frac n2 -
S_x\Bigr) }
where
\numeq{e19}{
S_x  = \fract12 \sum_{i=1}^n \sigma_x^{(i)}\ .
}
Choosing a linear interpolation between $\tilde H(0)$ and $H_P$ gives
\numeq{e20}{
\tilde H(s) = (1-s) \tilde H(0) + s H_P \text{\quad for $0\le s\le 1$.}
}
Since $\tilde H(s)$ depends only on the two operators $S_x$ and $S_z$, we start
our analysis by considering the states $\ket\theta$ defined by
\numeq{e21}{
(\sin\theta S_x + \cos\theta S_z) \ket\theta = \frac n2 \ket\theta\ .
}
Note that
\begin{subequations}\label{e22}
\begin{align}
\bra\theta S_x \ket\theta &= \frac n2 \sin\theta\\
\bra\theta S_x^2 \ket\theta &= \Bigl(\frac n2 \sin\theta\Bigr)^2 + O(n)\\
\bra\theta S_z \ket\theta &= \frac n2 \cos\theta\\
\bra\theta S_z^2 \ket\theta &= \Bigl(\frac n2 \cos\theta\Bigr)^2 + O(n)\\
\bra\theta S_z^3 \ket\theta &= \Bigl(\frac n2 \cos\theta\Bigr)^3 + O(n^2)
\end{align}
\end{subequations}
Now define
\numeq{e24}{
V(\theta,s) = 2(1-s) (1-\sin\theta) + sg\bigl(\fract12 (1-\cos\theta)\bigr)
}
where $g$ is given in \refeq{e15}. Using \refeq{e16}, \refeq{e18}, \refeq{e20}, and
\refeq{e22} we have
\numeq{e23}{
V(\theta,s) = \Bigl(\frac2n\Bigr)^3 \bra\theta \tilde H(s) \ket\theta + 
O(1/n)
}
where we have also used the operator relation $w=\frac n2- S_z$.  The function
$V(\theta,s)$, for each~$s$, represents a large-$n$ ``effective potential''. For  $n$
large, for each~$s$, the ground state of
$\tilde H(s)$ is well approximated by $\ket{\theta_m}$ where $\theta_m$
minimizes $V(\theta,s)$. 

\begin{figure}[ht]
\centerline{\BoxedEPSF{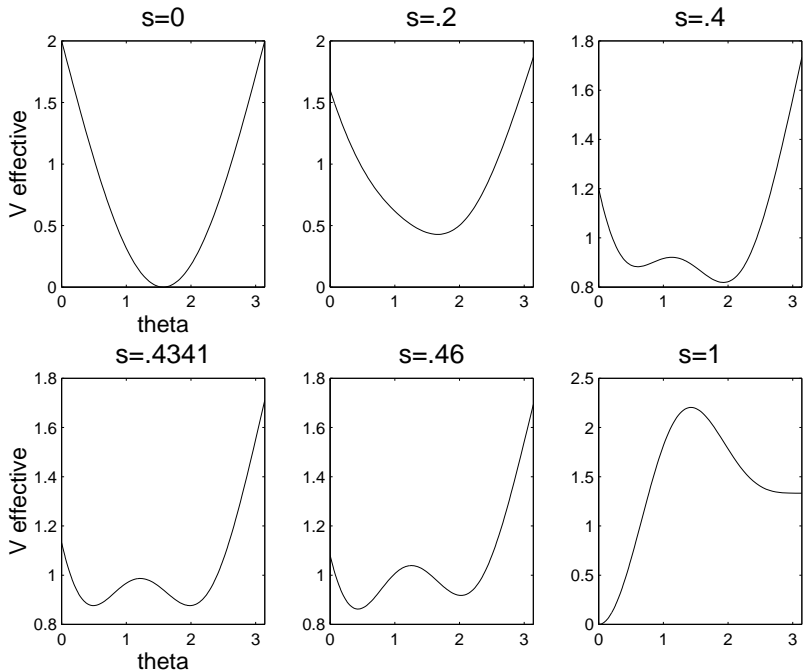 scaled 625}}
\caption{The effective potential versus $\theta$ changes shape as $s$ increases.}
\label{f2}
\end{figure}

In \refFig{f2} we plot $V(\theta,s)$ versus $\theta$ for $s=0,0.2,0.4,0.4341,0.46$,
and~1.  At $s=0$ the minimum is at $\theta=\pi/2$ corresponding to
Hamming weight $n/2$. Because $g'(1/2) < 0$, this minimum moves to
$\theta$ bigger than $\pi/2$ for small~$s$ as can be seen in the $s=0.2$
panel.  As~$s$ increases, a local minimum forms at values of~$\theta$ less
than
$\pi/2$. There is a critical value of~$s$, $s^* = 0.4341$, at which the two
minima occurring at $\theta_1 <\pi/2$ and at $\theta_2>\pi/2$ are
degenerate; that is,
$V(\theta_1, s^*) = V(\theta_2, s^*)$. 
For $s>s^*$ the global minimum is always at
a value of $\theta$ less than $\pi/2$. For $s=1$ this global minimum is at
$\theta_1=0$. 

For the \qaea\ to work, the quantum state of the system should remain in (or
very near) the instantaneous ground state of $\tilde H(s)$ with $s=t/T$. For
$s<s^*$ the ground state is smoothly varying. However, at $s=s^*$ the ground
state changes from $\approx \ket{\theta_2}$ to $\approx \ket{\theta_1}$. For
the quantum system to go from $\ket{\theta_2}$ to $\ket{\theta_1}$ requires
quantum tunneling through the barrier separating the minima at $\theta_1$
and $\theta_2$. This takes a time exponential in~$n$ and implies that the gap of 
$\tilde H(s^*)$ is exponentially small. Using a combination of standard large spin
and instanton methods\citer{r4} we can estimate the dominant exponentially small
term. We obtain
\numeq{e25}{
\text{gap} \sim \text{poly}(n) e^{-S_0 n}
}
where
\begin{subequations}\label{e26}
\begin{gather}
S_0 = \fract12 \int_{\theta_1}^{\theta_2}\!\! \phi(\theta) \sin\theta \rd\theta
\label{e26a}\\
\intertext{and}
\phi(\theta) = \cosh^{-1} \Bigl[1+\frac{V(\theta,s^*) -
V(\theta_1,s^*)}{2(1-s^*)\sin\theta}\Bigr]\ .
\label{e26b}
\end{gather}
\end{subequations}

\begin{figure}[tb]
\centerline{\BoxedEPSF{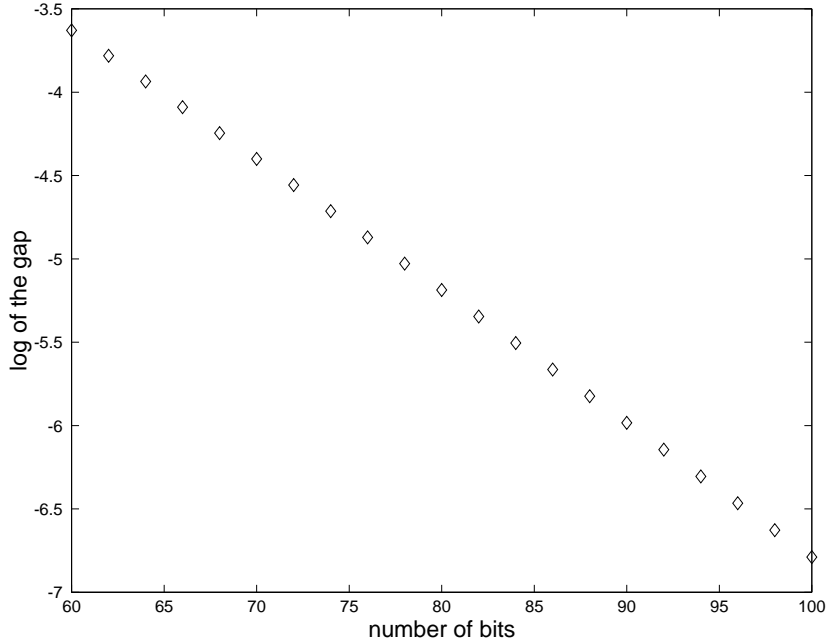 scaled 625}}
\caption{The log of the gap plotted versus~$n$. The straight line behavior
indicates that the gap is decreasing  exponentially in~$n$.}
\label{f3}
\end{figure}

For $q=3$, we find $s^*= 0.4341$ and $S_0 = 0.2015$. We can also determine the
minimum gap of $\tilde H(s)$ by numerical  diagonalization. Because of the
symmetry of $\tilde H$ and the initial state, although the full Hilbert space is
$2^n$ dimensional, the evolution takes place in the ($n+1$)-dimensional subspace
of total spin~$n/2$.  In \refFig{f3} we show the log of the gap versus~$n$ for
$n$~between 60 and~100.
At $n=100$ the minimum gap occurs at $s=0.4338$. Fitting the gap to the
functional form $An^p e^{-cn}$ and finding the best values of $A$, $p$, and~$c$
gives $c=0.2021$, which certainly supports our approach. For $q=5$ we have
$S_0=0.4946$ and $c=0.4966$, and for $q=7$ we have $S_0=0.7008$ and
$c=0.6930$.

Formula \refeq{e25} is valid for any example where the cost function depends
only on the Hamming weight~$w$ and we can write
\numeq{e27}{
h(w) = n^r g(w/n) + O(n^{r-1})
}
where the global minimum of $g(u)$ occurs at $u<1/2$ and $g'(1/2)<0$.
This guarantees that there is a value $s^*$ such that $V(\theta,s^*)$ given by
\refeq{e24} has degenerate minima at $\theta_1<\pi/2$ and $\theta_2>\pi/2$.
Then \refeq{e26} can be used to calculate the coefficient of $n$ in the exponent
of~\refeq{e25}. 

We now study the performance of simulated annealing for this kind of problem.
Again we start with a cost function $h(\znn)$, which is a sum of local terms but
because of symmetry only depends on the Hamming weight, so we can write the
cost function as $h(w)$. Now for any temperature~$\tau$, the \Bd\ \refeq{e8}
depends only on the Hamming weight of the string $w= \zsn$. The number of
strings with Hamming weight $w$ is $(\begin{smallmatrix}
n\\w\end{smallmatrix})$. We can define the (unnormalized) probability
\numeq{e28}{
p_\tau(w) = \Bigl(\begin{matrix}
n\\w\end{matrix}\Bigr) \exp \bigl(-h(w)/\tau\bigr)
}
which gives the probability of finding a string with Hamming weight~$w$ at
temperature~$\tau$ in the \Bd.

We also define the ``Free energy'', $F(w,\tau)$, by 
\numeq{e29}{
\exp\bigl(-F(w,\tau)/\tau\bigr) = \Bigl(\begin{matrix}
n\\w\end{matrix}\Bigr)  \exp\bigl(-h(w)/\tau\bigr) \ .
}
Note that for any~$\tau$, the most likely value of the Hamming weight occurs at
the value of $w$ that minimizes $F(w,\tau)$. Now we further specialize to the
case \refeq{e13} so that \refeq{e14} and \refeq{e15} apply. If we rescale the
temperature and let $a=8\tau/n^2$ then we can write 
\numeq{e30}{
F(w, an^2/8) = \Bigl(\frac n2\Bigr)^3 \mathcal{F}(u,a) +  O(n^2)
} 
where 
\numeq{e31}{
\mathcal{F}(u,a) = a\bigl[ (1-u)\log (1-u) + u\log u  \bigr] + g(u)
}
and again $u=w/n$.

\begin{figure}[tb]
\centerline{\BoxedEPSF{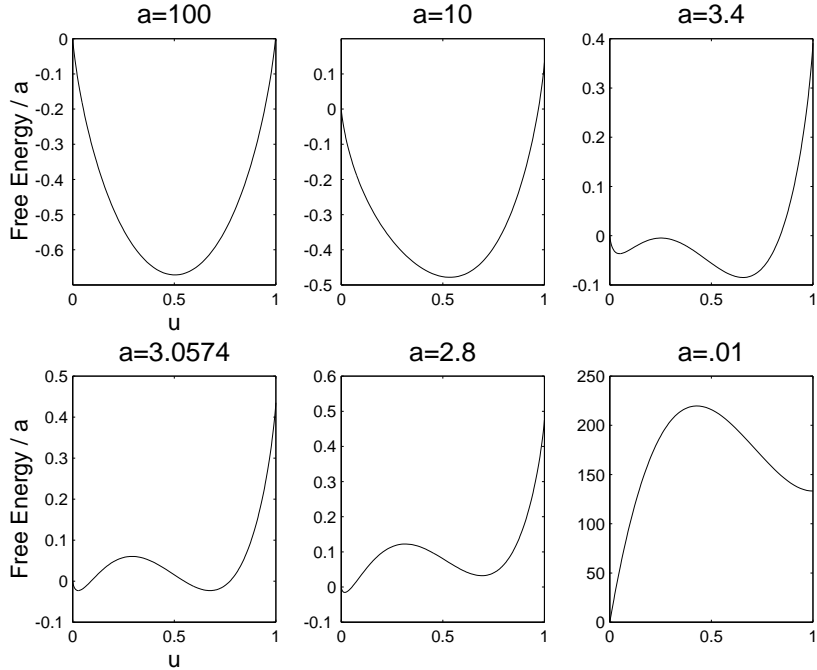 scaled 625}}
\caption{The scaled free energy $\mathcal F/a$ plotted as a function of the scaled
Hamming weight changes shape as the temperature decreases. Compare to
\protect\refFig{f2}.}
\label{f4}
\end{figure}

For $n$ large, for any value of the temperature $\tau=an^2/8$, the \Bd\ is
peaked on those strings with Hamming weight $nu$ where $u$ minimizes
\refeq{e31}. At infinite temperature, the first term in \refeq{e31} dominates
and is minimized at $u=1/2$. As $a$ is decreased the minimum moves to a
value of~$u$ above~1/2.  In \refFig{f4} we show $\mathcal F(u,a)/a$ versus $u$
for $a=100,10,3.4,3.0574,2.8$, and 0.01. The similarity with \refFig{f2} is
apparent. The critical value of~$a$, $a^* = 3.0574$, is where the global minimum
switches from a value of~$u$ above 1/2 to a value below.
To follow the \Bd, in the simulated annealing Markov chain, the Hamming weight,
$nu$, must change from the value corresponding to the larger local minimum to the
smaller while~$\tau$ is approximately~$a^* n^2/8$.

This order-$n$ change in the Hamming weight requires exponentially many steps:
The sign of  the derivative
$\frac\partial{\partial u} \mathcal F (u,a^*)$ determines whether the Markov
chain is more likely to move left or right. For large~$n$, if in the steps of the
annealing process the bits are chosen uniformly at random, $w$ is more
likely to move to $w+1$ than to $w-1$ if $\frac\partial{\partial u} \mathcal F
(u,a^*)< 0$. This means that climbing the hill separating the local minima is
exponentially unlikely in polynomially many steps. 

As long as $g(u)$ has a global minimum at $u<1/2$ and $g'(1/2)<0$ simulated
annealing, like quantum adiabatic evolution, will require a time exponential in $n$
to succeed. The method used in  this section, tracking the local minima of the
effective potential and the free energy, can be used to show the similarity of  the
performance of \qae\ and simulated annealing for many examples with a cost that
depends on the Hamming weight alone.

\subsection{The Hamming weight with a spike}
\label{sec5}

Here we consider the cost function
\numeq{e32n}{
h (w) = \begin{cases}
w, & w\neq n/4\\
n, & w = n/4\  .
\end{cases}
} 
where again $w$ is the Hamming weight of $n$~bits and also $n$ is taken to be a
multiple of~4. Note that this cost function cannot be written as a sum of terms
each of which involves only a few bits. The global minimum of $h(w)$ is at $w=0$
and there is  a local minimum at $w=(n/4) + 1$.

It is easy to see that simulated annealing fails to find the global minimum in
polynomial time. For $\tau> 1/\log3$, the \Bd\ concentrates at Hamming weights
larger than $n/4$. For $\tau<1/\log3$ the \Bd\ concentrates at Hamming weights
smaller than $n/4$. At temperatures $\sim1/\log3$, according to the acceptance
rule~\refeq{e10}, a string with Hamming weight $(n/4)+1$ will only flip to a
string with Hamming weight $n/4$ with a probability that is exponentially small
in~$n$. Thus a simulated annealing algorithm running for only polynomial time
gets hung up in the false minimum of~$h$. 

\begin{figure}[ht]
\centerline{\BoxedEPSF{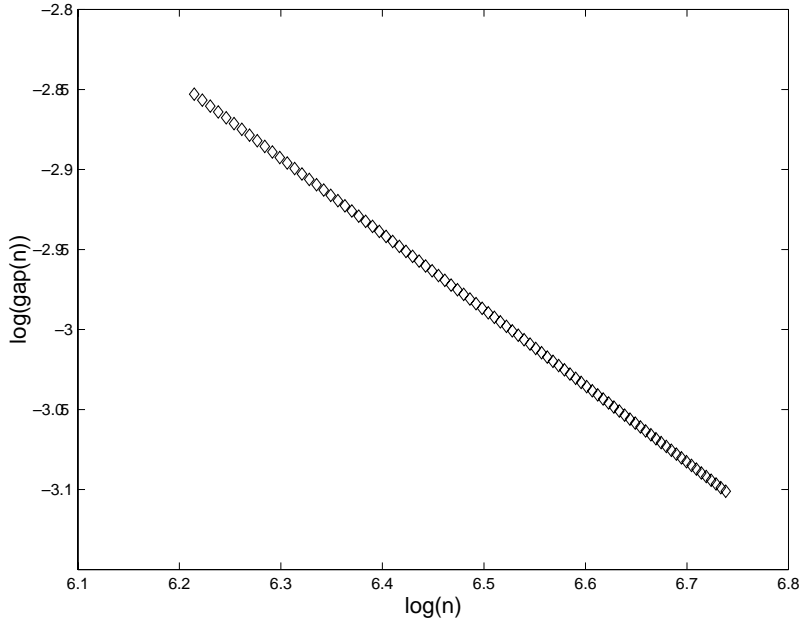 scaled 600}}
\caption{The Hamming weight with a spike. The straight line behavior on the
log-log plot indicates that the gap decreases polynomially in~$n$.}
\label{newfig}
\end{figure} 

In quantum mechanics, it is well known that
a low-energy particle can penetrate a high potential barrier. If the barrier is thin,
the penetration probability need not be small. This turns out to be the case here.
As in the examples of Section~\ref{sec4}, since the cost is a function only of the
Hamming weight, the quantum evolution takes place in an ($n+1$)-dimensional
subspace. The gap can be calculated by explicit diagonalization (using a technique
similar to the one in\citer{r1}, section~4.2). The result is that
\numeq{e32nn}{
\mathrm{gap}(n) \approx 1.35\, n^{-1/2}
}
and, for $n$ large, occurs at $s=0.366$. We can also find the gap by numerical
diagonalization; see \refFig{newfig}, which plots $\log \mathrm{gap}(n)$ versus
$\log (n)$ for $500\le n\le848$. The straight line fit has slope $-0.474$,
consistent with the predicted $-0.5$ in~\refeq{e32nn}. For $n=848$, the gap
occurs at 0.3675, close to the predicted 0.366.  

\subsection{The bush of implications}
\label{sec6}
 
In the previous section we discussed an example with a cost function that treats
all bits symmetrically and depends only on the Hamming weight of the $n$ bits.
Here we describe an example where the cost function depends on the Hamming
weight of $n$ bits, but also on the value of a single additional bit. As in
Section~\ref{sec4}, we consider the effective potential and free energy.
In the example in this section, \qae\ succeeds in polynomial time while simulated
annealing does not. 
\begin{figure}[ht]
\centerline{\BoxedEPSF{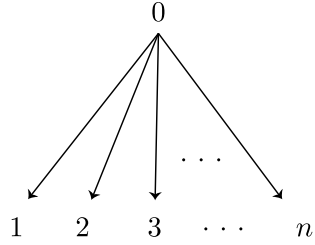 scaled 1100}}
\caption{The bush of implications.}
\label{f5}
\end{figure} 

We label the $n+1$ bits $0,1,2,\ldots,n$. The cost function is given by
\numeq{e5.1}{
h (z_0,\znn) = \sum_{i=1}^n z_0 (1-z_i) + (1-z_0)\ .
}
The terms $z_0 (1-z_i) $ can be viewed as ``imply'' clauses, since they are~0
unless $z_0=1$ and $z_i = 0$. 
The unique string with $h=0$ has $z_0=1$ and all $z_i = 1$. If $z_0=0$, then
$h=1$ no matter what the values of $\znn$ are. If $z_0=1$, then $h$ equals $n-w$
where $w= \zsn$ is the Hamming weight of the string $\zn$. We can rewrite
\refeq{e5.1} as 
\numeq{e5.2}{
h(z_0,w) = z_0 (n-w) + 1 - z_0
}
and as before
\numeq{e5.3}{
h(z_0,w) = ng \bigl(z_0, \frac wn\bigr) + O(1)
}
with
\begin{subequations}\label{e5.4}
\begin{align}
g(0,u) &= 0\label{e5.4a}\\
\intertext{and}
g(1,u) &= 1-u\ .\label{e5.4b}
\end{align}
\end{subequations}
Note the degeneracy of the minima of $g$; this is  removed by including the lower
order term $(1-z_0)$. 

\begin{figure}[ht]
\centerline{\BoxedEPSF{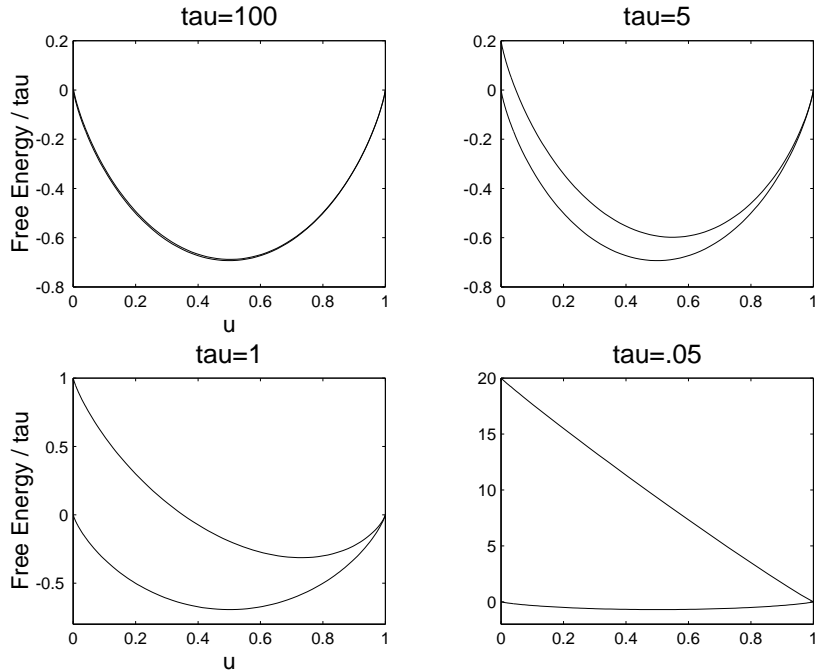 scaled 625}}
\caption{The bush of implications. The scaled free energy plotted versus the scaled
Hamming weight as the temperature changes. In each panel the top curve
corresponds to
$z_0=1$ and the lower corresponds to $z_0=0$. The global minimum never moves
from
$z_0=0, u=1/2$.}
\label{f6}
\end{figure}

As before, to determine the performance of annealing, we calculate the free
energy $\mathcal F$, defined by
\numeq{e5.5}{
\exp\bigl(- F_{z_0} (w,t)/\tau\bigr) = \Bigl(\begin{matrix}
n\\w\end{matrix}\Bigr) \exp \bigl(-h(z_0,w)/\tau\bigr)
}
and
\numeq{e5.6}{
F_{z_0} (w,\tau) = n \mathcal F_{z_0} \Bigl(\frac wn , \tau\Bigr) + O(1)\ .
}
We get
\begin{subequations}\label{e5.7}
\numeq{e5.7a}{
\mathcal F_{0} (u,\tau) = \tau \bigl[ (1-u)\log (1-u) + u\log u  \bigr]
} 
and 
\numeq{e5.7b}{
\mathcal F_{1} (u,\tau) = \tau \bigl[ (1-u)\log (1-u) + u\log u  \bigr] + (1-u)\ .
} 
\end{subequations}
In \refFig{f6} we plot $\mathcal F_{z_0} (u,\tau) /\tau$, $\tau=100,5,1$,
and~0.05.  We see that with the $O(1)$ terms neglected, the value $z_0=0$,
$w=\frac n2$ minimizes the free energy for all~$\tau$, $0<\tau<\infty$. 

Can the addition of the
correction, which adds~1 to $F_0 (w,\tau)$ for all $(w,\tau)$, make any
difference in a simulated annealing process? For $\tau=0$, the $1-z_0$ term
means that the global minimum of~$h$ occurs at $z_0=1$, $w=n$. 
For $\tau\approx1$, a simulated annealing process that is following the \Bd\
will be concentrated at $z_0=0$, $w=\frac n2$.  There are two routes from
$z_0=0$, $w=\frac n2$  to $z_0=1$, $w=n$. One route is to jump from $z_0=0$ to
$z_0=1$ while $w\approx\frac n2$. But the probability of accepting this change
is exponentially small in $n$ for $\tau<1$, so this route is exponentially unlikely
in polynomially many steps. (Note that changing the rate at which bit~0 is
flipped does not help.)
The other route is to maintain $z_0=0$ while $w$ increases. But this is exactly
the kind of hill climbing in free energy [this time $\mathcal F_0(u,\tau)$] that
cannot occur in polynomial time as in the example in the previous section. In
conclusion, simulated annealing fails to reach the true minimum of~$h$ at
$z_0=1$, $w=n$ in polynomial time.

We now turn to the behavior of a \qaa\ for the bush of implications. Using spin
operators we write the quantum Hamiltonian corresponding to $h$ of 
\refeq{e5.1} as
\numeq{e5.8}{
H_P = \fract12 (1-\sigma_z^{(0)}) \Bigl(\frac n2 + S_z\Bigr) +
\fract12 (1+\sigma_z^{(0)})
}
where $S_z$ is the $z$~component of the total spin for bits 1 through~$n$,
whereas $\fract12  \sigma_z^{(0)}$ is the spin component of the spin of bit~0.
Now bits 1 through~$n$ are each involved in one term in \refeq{e5.1} whereas
bit~0 is in $n+1$ terms. Accordingly, we get for $\tilde H (0)$
[see\citer{r1}, formula (2.22)]
\numeq{e5.9}{
\tilde H (0) = (n+1) \fract12 (1-\sigma_x^{(0)}) + \Bigl(\frac n2 - S_x\Bigr)\ .
}
For the interpolating Hamiltonian we take
\numeq{e5.10}{
\tilde H (s) = (1-s) \tilde H(0) + s H_P\ .
}

Consider the states $\ket{\alpha,\theta} = \ket\alpha \otimes \ket\theta$ where
$\ket\alpha$ is the bit~0 state
\numeq{e5.11}{
\ket\alpha = (\sin\alpha/2) \ket{z_0=0} +
(\cos\alpha/2) \ket{z_0=1}
}
and $\ket\theta$ is made of bits 1 through~$n$ and is  defined by \refeq{e21}.
For the effective potential we set
\numeq{e5.13}{
V(\alpha,\theta,s) =  (1-s) [2-\sin\alpha-\sin\theta] + 2s\cos^2\frac\alpha2 
\cos^2\frac\theta2
}
which  gives
\numeq{e5.12}{
V(\alpha,\theta,s) = \frac2n \bra{\alpha,\theta} \tilde H(s) \ket{\alpha,\theta}
+ O\Bigl(\frac1{n^2}\Bigr)\ . 
}
Note that $V(\alpha,\theta,s)=V(\theta,\alpha,s)$ for all~$s$.
\begin{figure}[tb]
\centerline{\BoxedEPSF{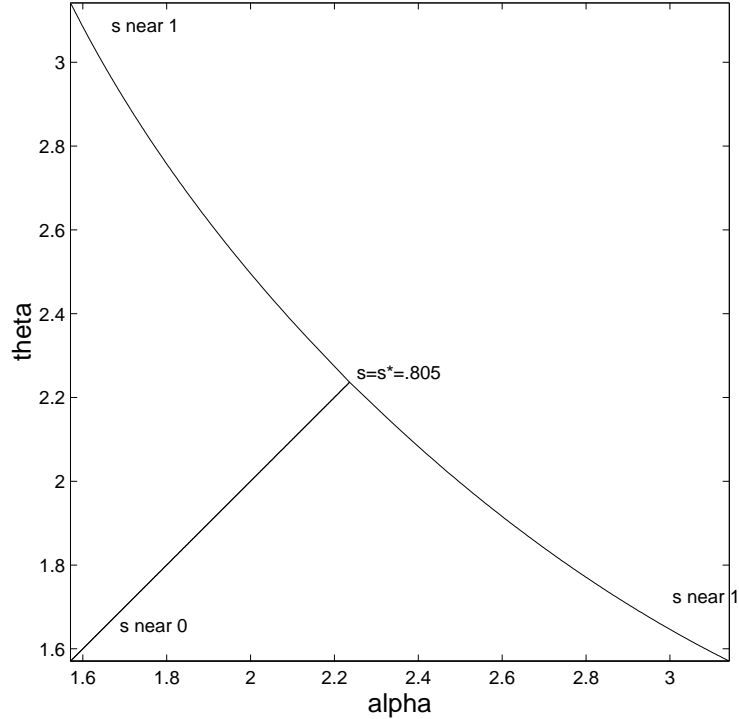 scaled 705}}
\caption{The bush of implications. The global minimum of the effective potential 
plotted as a function of~$s$.}
\label{f7}
\end{figure}

In \refFig{f7} we plot $\bigl(\alpha(s),\theta(s)\bigr)$, the global minimum of
$V(\alpha,\theta,s)$, as a parameterized curve for $0\le s\le1$.  At $s=0$ the
global minimum occurs at $\alpha=\pi/2$, $\theta=\pi/2$, which corresponds to
a uniform superposition for bit~0 and a Hamming weight concentrated at $n/2$
for bits~1 through~$n$. As $s$ increases the minimum is at 
$\alpha(s)=\theta(s)$ until a critical value $s=s^*=0.805$. For $s>s^*$ there are
two global minima symmetric under interchange of $\alpha$ and~$\theta$. 
We go beyond the effective potential and analyze the quantum corrections for
$s=s^*$, using a representation in terms of a 2-component (to represent the state
of bit~0) anharmonic oscillator (to represent the fluctuations about the state
$\ket{\theta(s^*)}$). We find an expansion in powers of $n^{-1/3}$. The symmetry
under interchange of $\alpha$ and $\theta$ is broken by the quantum corrections,
and the branch heading towards 
$\alpha=\pi/2$, $\theta=\pi$ has the lower energy. The adiabatic evolution is
therefore heading towards the state  that is a uniform superposition for bit~0 and
has Hamming weight~$n$ for bits~1 through~$n$. This analysis so far has
neglected the $\frac12 (1+\sigma_z^{(0)})$ term in $H_P$ because it is lower
order in~$n$. For $(1-s)$ of order~$1/n$, the effect of this term in $\tilde H(s)$
is of the same order as $\tilde H(0)$. The net effect of this term is to rotate bit~0
to $\ket{z_0=1}$ in an $n$-independent fashion.

According to this analysis the minimum gap appears at $s=s^*+O(n^{-2/3})$,
where
$s^*$ is the bifurcation point in \refFig{f7}. 
We calculate 
\numeq{e5.14}{
\text{gap}(n) = \frac{0.5782}{n^{1/3}} + \frac{B}{n^{2/3}}  
+ O(n^{-1}) \ .
}
\begin{figure}[tb]
\centerline{\BoxedEPSF{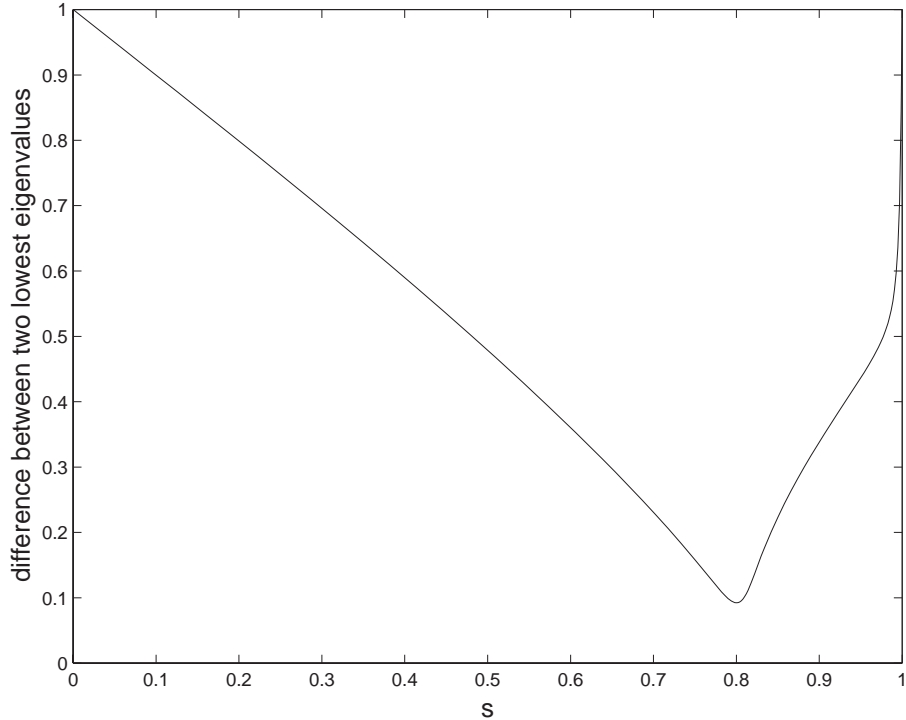 scaled 700}}
\caption{The bush of implications. The difference between the two lowest
eigenvalues of the Hamiltonian as a function of~$s$, when $n=300$. Note that the
minimum occurs near the predicted point $s^*=0.805$.}
\label{f8}
\end{figure}%

\begin{figure}[tb]
\centerline{\BoxedEPSF{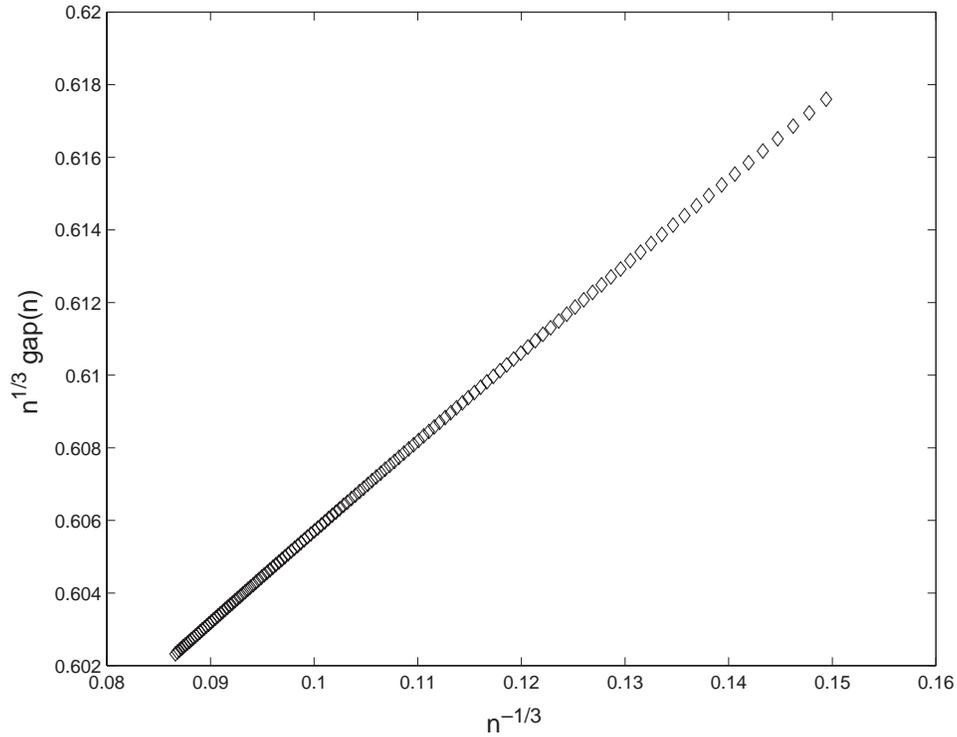 scaled 700}}
\caption{The bush of implications. The gap multiplied by $n^{1/3}$ and plotted
versus $n^{-1/3}$. The intercept of a straight line fit is 0.5812, compared to the
predicted 0.5782.}
\label{f9}
\end{figure}%

We can confirm this analysis numerically because the Hamiltonian $\tilde H(s)$
acts in a reduced Hilbert  space of dimension $2(n+1)$. In \refFig{f8} we show the
difference between the  lowest two eigenvalues $E_1(s) -E_0(s)$ for $0\le s\le1$ at
$n=300$. Note that the minimum difference occurs at $s=0.800$. In \refFig{f9} we
plot $n^{1/3}\cdot\mathrm{gap}(n)$ versus $n^{-1/3}$ for $300\le n \le1540$.
According to prediction \refeq{e5.14} this should  be a straight line (with
corrections of order~$n^{-2/3}$) with an intercept of 0.5782. A straight-line fit to
the data in
\refFig{f9} has an intercept of 0.5812. 

It is worth noting that the choice of $\tilde H(0)$ in \refeq{e5.9} was made
so that $\tilde H(s)$ would be the sum of terms
\numeq{e48}{
\tilde H(s) = \sum_C \tilde H_C (s)
}
where each $ \tilde H_C (s)$ depends only on clause~$C$ (see\citer{r1},
formula~2.28). Suppose instead, for the bush of implications, we write
\numeq{e49}{
\tilde H (0) = \lambda (n+1) \fract12 (1-\sigma_x^{(0)}) + \Bigl(\frac n2 -
S_x\Bigr)
 }
with $\lambda$ a parameter. We can recalculate the effective potential and draw
the picture analogous to \refFig{f7}.  If $\lambda>1$, the gap for $n$~large is a
positive constant. If $\lambda<1$, the gap is exponentially small. Note that the
choice $\lambda = 1/(n+1)$ corresponds to a magnetic field in the $x$-direction
with the same strength for each bit. This case, mentioned in\citer{r1}, is an
example of a local problem with a choice for a \qaa\ that requires exponential
time. This illustrates that the cost function alone does not determine the
performance of the \qa\ but rather that the choice of the path to $H_P$ is
sometimes crucial.

\subsection{Conclusion}
\label{sec7}

In this paper we first compared the behavior of \qae\ and classical simulated
annealing in searching for the global minima of cost functions that depend on
$n$~bits symmetrically. When the cost depends smoothly on the Hamming weight
of the~$n$ bits, \qae\ and simulated annealing typically perform similarly.

We have also shown, however, that there are examples where this similarity
breaks down. In the spike example, the cost function has a barrier that can be
penetrated quantum mechanically but not classically, so the \qa\ succeeds in
polynomial time whereas annealing does not. In the bush example, the addition of
a single spin-1/2 leads to quantum behavior with no classical analogue.  These
examples suffice to show that there is no theorem that says \qae\ will fail when
simulated annealing fails. It remains an open question whether they also indicate
that \qaea s will succeed on problems of computational interest that cannot be
solved efficiently by classical local search.

\subsection*{Acknowledgments}
This work was supported in part by the Department of Energy under cooperative
agreement DE--FC02--94ER40818 and by the National Security Agency (NSA) and
Advanced Research and Development Activity (ARDA) under Army Research
Office (ARO) contract DAAD19-01-1-0656.  We  thank Andrew Childs and John
Preskill for helpful discussions and Umesh Vazirani for sharing some of his group's
results prior to posting. We also thank Martin Stock for his help preparing this
manuscript.

\end{document}